\documentclass[draft]{agujournal2019}
\usepackage{url}
\usepackage{lineno}
\usepackage[inline]{trackchanges}
\usepackage{soul}
\usepackage{amsmath}
\usepackage{natbib}
\usepackage{lmodern}

\usepackage{xcolor}
\usepackage{multirow}

\draftfalse

\journalname{Journal of Geophysical Research: Space Physics}
\newcommand{\solphys}{Solar Physics}

\begin{document}

\title{Daily Predictions of F10.7 and F30 Solar Indices with Deep Learning}

\authors{Zhenduo Wang\affil{1,2}, 
Yasser Abduallah\affil{1,2},
Jason T. L. Wang\affil{1,2}, Haimin Wang\affil{1,3,4}, 
Yan Xu\affil{1,3,4},
Vasyl Yurchyshyn\affil{4},
Vincent Oria\affil{1,2}, 
Khalid A. Alobaid\affil{5}, 
Xiaoli Bai\affil{6}}
\affiliation{1}{Institute for Space Weather Sciences, New Jersey Institute of Technology, 
Newark, NJ 07102, USA}
\affiliation{2}{Department of Computer Science, New Jersey Institute of Technology, 
Newark, NJ 07102, USA}
\affiliation{3}{Center for Solar-Terrestrial Research, New Jersey Institute of Technology,  
Newark, NJ 07102, USA}
\affiliation{4}{Big Bear Solar Observatory, New Jersey Institute of Technology, 
Big Bear City, CA 92314, USA}
\affiliation{5}{College of Applied Computer Sciences, King Saud University, Riyadh 11451, Saudi Arabia}
\affiliation{6}{Department of Mechanical and Aerospace Engineering, Rutgers University,
Piscataway, NJ 08854, USA}


\begin{keypoints}
\item 
We present a deep learning model, named SINet, 
for making daily predictions of F10.7 and F30 
solar indices (1-60 days in advance) 
\item 
SINet outperforms 
five closely related statistical and deep learning methods
for F10.7 prediction

\item 
SINet is the first deep learning method for F30 prediction
\end{keypoints}

\begin{abstract}
The F10.7 and F30 solar indices are the solar radio fluxes measured at wavelengths
of 10.7 cm and 30 cm, respectively, which are key indicators of solar activity. 
F10.7 is valuable for explaining the impact of 
solar ultraviolet (UV) radiation on the upper atmosphere of Earth, 
while F30 is more sensitive and could improve the reaction of thermospheric density to solar stimulation. 
In this study, we present a new deep learning model, named the
Solar Index Network, or SINet for short,
to predict daily values of the F10.7 and F30 solar indices.
The SINet model is designed to make
medium-term
predictions of the index values
(1-60 days in advance).
The observed data used for SINet training were taken from the National Oceanic and Atmospheric Administration (NOAA) as well as Toyokawa and Nobeyama facilities. 
Our experimental results show that
SINet performs better than 
five closely related statistical and deep learning methods
for the prediction of F10.7. 
Furthermore, to our knowledge, this is the first time deep learning has been used
to predict the F30 solar index.
\end{abstract}

\section*{Plain Language Summary}
F10.7 and F30 are key indices that reflect the level of solar activity. F10.7 is valuable for explaining the impact of 
solar ultraviolet (UV) radiation on the upper atmosphere of Earth, 
while F30 is more sensitive and could improve the reaction of thermospheric density to solar stimulation.
We introduce a new forecasting model, named SINet, to
make medium-term daily predictions of the F10.7 and F30 solar indices.
The forecast horizons range from 1 to 60 days.
Our experimental results 
show that SINet achieves an
average mean absolute percentage error (MAPE) of
10.1\% for F10.7 and 9.5\% for F30 at the $60^{th}$ day of forecast,
better than the average MAPE of 
10.4\% for F10.7 and 9.6\% for F30 obtained by
the best one (temporal convolutional network)
of five closely related methods.

\section{Introduction}

The F10.7 and F30 solar indices refer to the solar radio fluxes measured at wavelengths of 10.7 cm and 30 cm, respectively.
They are critical gauges of solar activity, reflecting the intensity of radio emissions from the Sun's corona \citep{2015JGRA..120.6779G,2021ApJS..254....9P,2024SoPh47Z}. 
These indices are measured in solar flux units (sfu),
where 1 sfu is equal to $10^{-22}Wm^{-2}Hz^{-1}$. 
They range from less than 50 sfu to more than 300 sfu.  
F10.7 is valuable for explaining the impact of 
solar ultraviolet (UV) radiation on the upper atmosphere of Earth, 
while F30 is more sensitive and could improve the reaction of thermospheric density to solar stimulation \citep{2017JSWSC...7A...9D}. 
Although F10.7 is commonly used in space weather
forecasting \citep{2013SpWea..11..394T}, some researchers
recommend using F30 as the optimal proxy for solar activity \citep{2023SpWea..2103359L}.

Many techniques have been
developed to forecast the F10.7 and F30 solar indices.
For example,
 \citet{2017SpWea..15.1039W} developed a linear forecasting model to predict F10.7.
  \citet{2019AdAst2019E..19L} constructed an empirical model to predict F10.7 using extreme ultraviolet (EUV) images.   
\citet{2021ApJS..254....9P} utilized the adaptive Kalman filter to improve the McNish-Lincoln method
to predict F10.7 and F30. 
These traditional methods could be used to capture some features of the F10.7 and F30 indices
but often rely on assumptions that limit their ability to capture complex, non-linear dynamics inherent in solar activity.

With the advent of deep learning, 
predictive models have advanced in forecast accuracy and model performance
\citep{2023NatSR..1313665A,2024SoPh..299..159A,
2022JGRA..12730868B,
2021JGRA..12628228T}. 
For example, 
\citet{2022SoPh..297..157Z} 
developed a long short-term memory (LSTM) network to
predict F10.7 with good results.
\citet{DBLP:journals/ascom/JerseM24} later 
designed an LSTM network with multihead attention mechanisms to predict F10.7.  
\cite{2023SpWea..2103675D} used neural network ensembles
to predict F10.7.
\citet{2024SpWea..2203552H} 
combined an LSTM network with
variational mode decomposition (VMD)
to make predictions of F10.7 (1 day in advance).
\citet{2024SoPh47Z} adopted the informer model to forecast F10.7 for up to 27 days.
More recently, 
\citet{2024AnGeo..42...91W} employed a 
deep temporal convolutional network (TCN)
to make predictions of the F10.7 index.
TCN has also been used to predict other solar events
\citep[see, e.g.,][]{2025ApJS..276...68X}.

In this study, we propose a new deep learning model, 
named the Solar Index Network, or SINet for short,
to make medium-term daily predictions of
the F10.7 and F30 solar indices (1-60 days in advance).
In general, short-term prediction has a forecast horizon of several days (up to one solar rotation,
that is, up to 30 days).
Medium-term prediction refers to predictions made
one to several months in advance. 
Long-term prediction is
made over a decade or more.
\citet{2021ApJS..254....9P} was able to use a traditional method, 
namely the adaptive Kalman filter,
to predict the F10.7 and F30 solar indices 24 months in advance.
However, the data they used for the predictions are
monthly mean data for the F10.7 and F30 indices.
The ratio between the largest forecast horizon and the cadence of the data
is 24:1.
\citet{2024SoPh47Z} used a deep learning method to make daily predictions of F10.7
(1-27 days in advance). 
The ratio between their largest forecast horizon and the cadence of the data is 27:1.
In contrast, our SINet model can make medium-term daily predictions of
F10.7 and F30 (1-60 days in advance).
The ratio between the largest forecast horizon and the cadence of the data is 60:1, 
far exceeding the ratios of the existing methods.
We attribute the ability to handle this large ratio to the deep architecture employed by SINet.

The remainder of this paper is organized as follows.
Section \ref{sec:data} describes the data collection and pre-processing for our work.
Section \ref{sec:method} presents the architecture and configuration
details of the SINet model.
Section \ref{sec:results} reports the experimental results.
Section \ref{sec:discussion} presents a discussion of the results.
Section \ref{sec:conclusion}
concludes the article.

\section{Data}
\label{sec:data}

\begin{figure}
\centering
\includegraphics[width=0.9\linewidth]{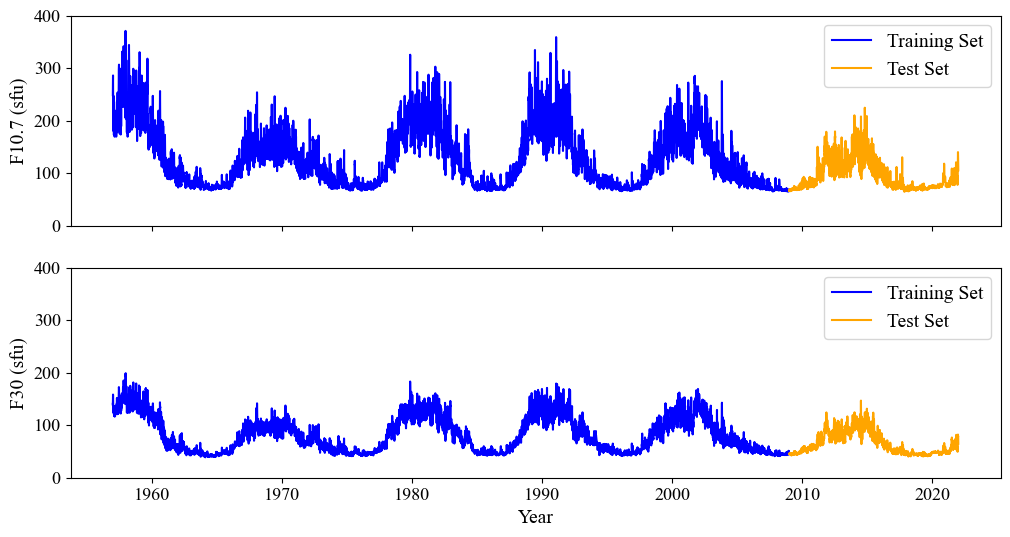}
\caption{Illustration of the time series data
sets for F10.7 (top) and F30 (bottom) used in our study.}
\label{fig:data}
\vspace*{-0.5cm}
\end{figure}

The daily values of the F10.7 solar index are obtained from the National Oceanic and Atmospheric Administration (NOAA) 
as in \citet{2024SpWea..2203552H}.
The daily values of the F30 solar index are collected from the Toyokawa and Nobeyama facilities 
as in \citet{2021ApJS..254....9P}.
These daily values form a one-dimensional (1D) time series
that spans 1957 to 2021.
The time series data are divided into training data from 1957 to 2008, 
10\% of which are used for validation, 
and test data from 2009 to 2021.
Figure \ref{fig:data} illustrates the time series data sets for F10.7 (top) and
F30 (bottom), respectively.
The training set and the test set for F10.7 (F30, respectively) are disjoint.
Thus, our SINet model is trained with data different from the test data and makes predictions on the test data that it has never seen during training.

Since the F10.7 data have different scales,  
we normalize the F10.7 data using the min-max method,
which is defined as follows:
\begin{equation}
X_{norm} = \frac{X - X_{min}}{X_{max} - X_{min}}.
\end{equation}
Here, $X$ represents a daily value of F10.7,
$X_{norm}$ represents the normalized value of $X$, 
$X_{min}$ is the minimum value of F10.7,
and $X_{max}$ is the maximum value of F10.7.
The min-max normalization method produces daily F10.7 values 
in the range between 0 and 1.
The F30 data are normalized similarly.

Data labeling is crucial for machine learning.
In the following, we describe how to label the F10.7 data.
The F30 data are labeled similarly, and its description is omitted.
We present two different approaches:
fixed prediction and
rolling prediction.
Consider each time point/day $d$ in the training set.
We create a training sample at the time point $d$ that contains
the historical F10.7 values for the previous 30 days.
That is, the training sample contains the true F10.7 values
on days
$d-30+1$, $d-30+2$,
$\ldots$,
$d-1$,
$d$.
For fixed prediction,
the labels of the training sample contain the
true F10.7 values on days 
$d+1$, 
$d+2$, 
$\ldots$, 
$d+59$,
$d+60$ (1-60 days in advance).
For rolling prediction, the label of the training sample is the true F10.7 value on day $d+1$ 
(1 day in advance). 

\begin{figure}[ht]
\centering
\includegraphics[width=0.9\linewidth]{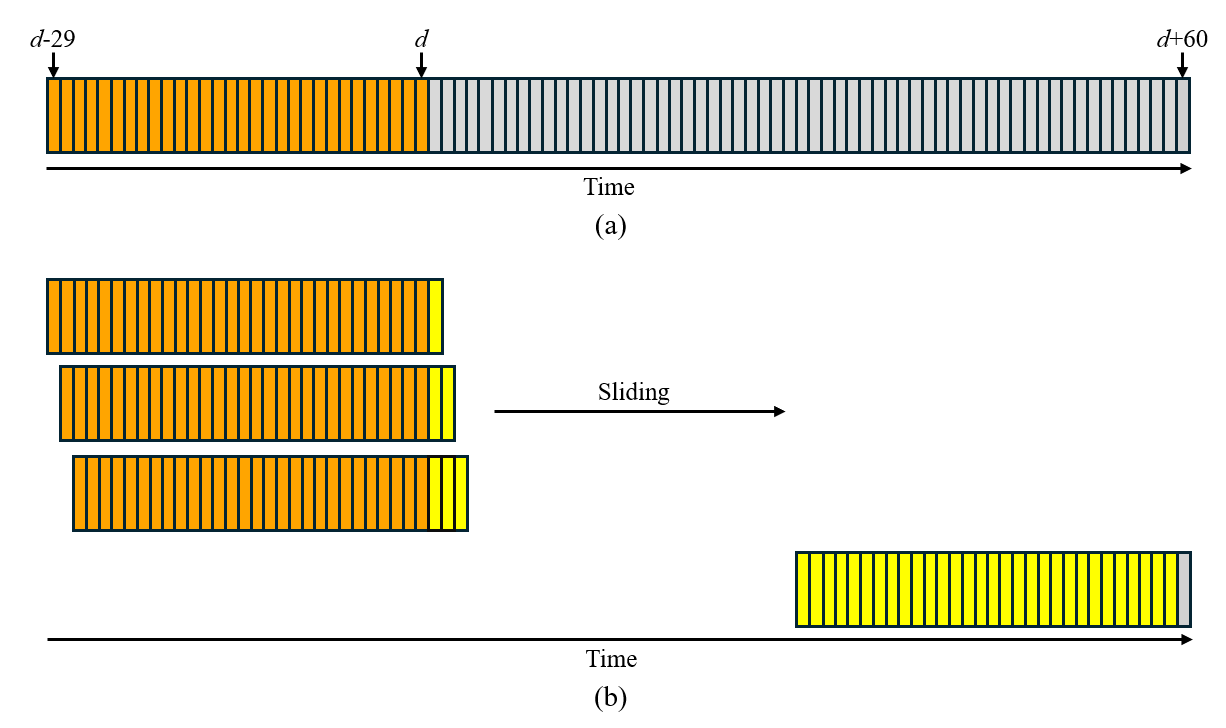}
\caption{Illustration of the two prediction approaches employed by SINet.
(a) The fixed prediction approach uses 
the historical F10.7 values of the previous 30 days,
$d-30+1$, 
$d-30+2$, 
$\ldots$, 
$d-1$, 
$d$.
represented by orange rectangles, to
predict the F10.7 values on days $d+1$, $d+2$, $\ldots$, $d+59$, $d+60$, represented by gray rectangles. 
(b) The rolling prediction approach uses a sliding window method, where each predicted value is appended iteratively to generate new predictions one day at a time, ultimately achieving a 60-day ahead forecast. Orange rectangles represent true historical F10.7 values, while yellow rectangles represent predicted synthetic F10.7 values 
that are used subsequently
to make new predictions. See text for detailed descriptions of the rolling prediction approach.}
\label{fig:fixrolling}
\vspace*{-0.5cm}
\end{figure}

Next, look at each time point/day $d$ in the test set.
We create a test sample at the time point $d$ that contains
the historical F10.7 values for the previous 30 days.
The labels of the test sample are absent and will be 
predicted by SINet.
Specifically, we use the historical F10.7 values
for the previous 30 days
to predict the F10.7 values (1-60 days in advance).
Figure \ref{fig:fixrolling}(a) illustrates the fixed prediction approach.
The 30 orange rectangles represent the historical F10.7 values of the previous 30 days:
$d-30+1$, 
$d-30+2$,
$\ldots$,
$d-1$,
$d$.
The 60 gray rectangles represent the predicted values of F10.7 on days 
$d+1$, 
$d+2$, 
$\ldots$,
$d+59$, 
$d+60$. 

Figure \ref{fig:fixrolling}(b) illustrates the rolling prediction approach, where a sliding window method is used.
We move to the right
one position/day at a time, and in each step we use the historical F10.7 values of the previous 30 days to predict the next F10.7 value (1 day in advance).
Initially, in step 1, we use the historical F10.7 values of the previous 30 days, represented by orange rectangles, to predict the F10.7 value 
(1 day in advance) on day $d+1$, represented by a 
yellow rectangle.
Then, in step $i$, $2 \leq i \leq 30$,
we use the historical F10.7 values of the previous $(30-i+1)$ days, represented by orange rectangles, combined with the $(i - 1)$ predicted synthetic values
obtained in the previous $(i - 1)$ steps, represented by yellow rectangles, to predict the F10.7 value (1 day in advance)
on day $d+i$, represented by a yellow rectangle.
In step $j$, $31 \leq j \leq 59$,
we use the synthetic F10.7 values obtained in step $(j-30)$ to step $(j-1)$
to predict the F10.7 value (1 day in advance) on day $d+j$.
In step $60$, 
we use the synthetic F10.7 values of the previous 30 days,
represented by yellow rectangles, obtained
in step 30 to step 59, to
predict the F10.7 value (1 day in advance) on day $d+60$,
represented by a gray rectangle.

\section{Methodology}
\label{sec:method}

\begin{figure}[t]
\centering
\includegraphics[width=\linewidth]{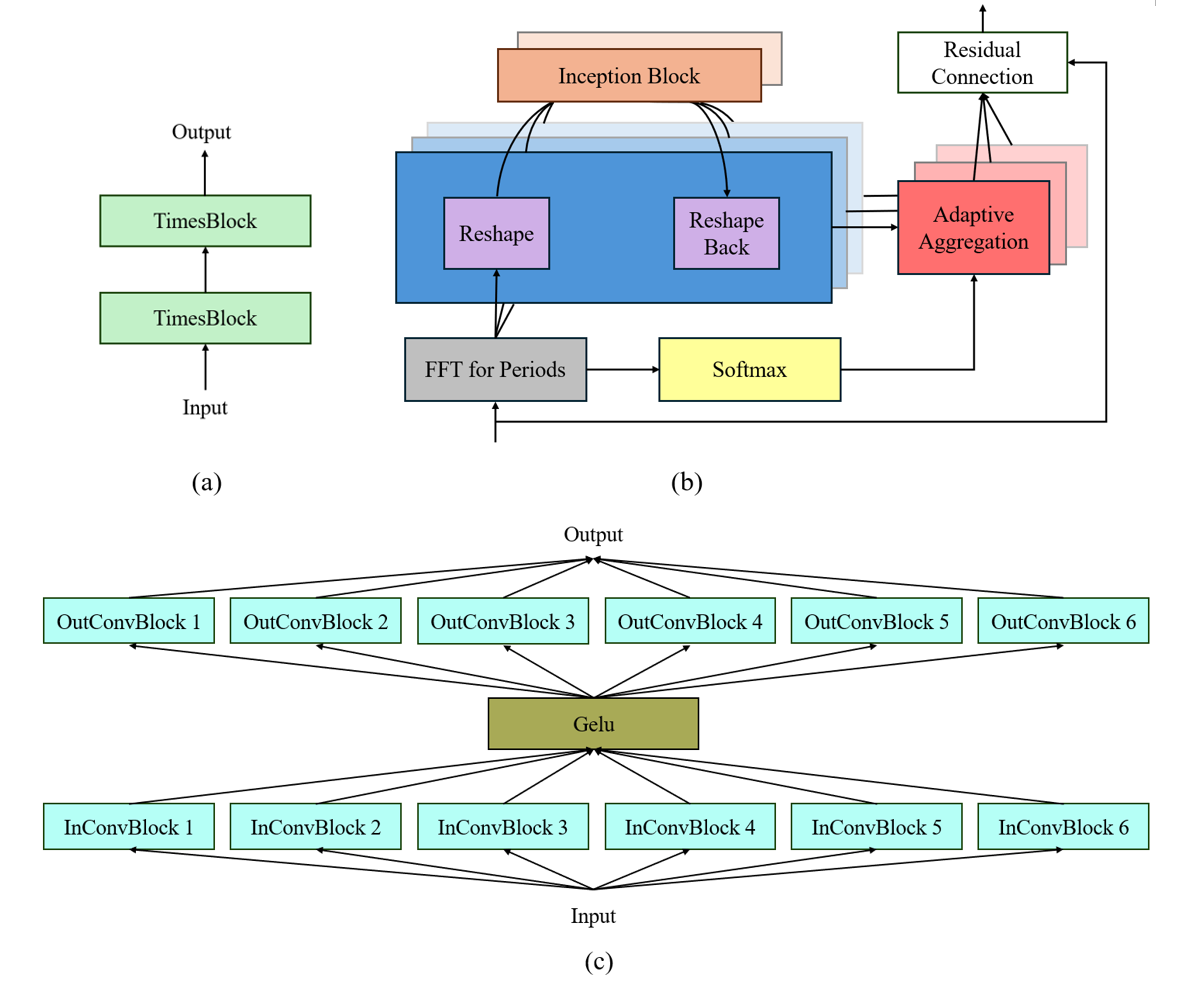}
\caption{Illustration of the SINet model architecture.
(a) Overall architecture of SINet, which contains two TimesBlocks.
(b) Architecture of a TimesBlock, 
which contains a dual-inception model structure with two inception blocks.
(c) Architecture of the dual-inception model structure, in which
the two inception blocks are connected by a Gelu layer.
See text for detailed descriptions of the components of SINet.}
\label{SINetarch}
\vspace*{-0.5cm}
\end{figure}

Figure \ref{SINetarch}(a) shows the architecture of SINet,
which is an enhancement of TimesNet \citep{DBLPWuHLZ0L23},
adapted for the prediction of solar indices. 
SINet employs two TimesBlocks.
Each TimesBlock is designed to process 1D time series and enhance feature representation.
During prediction/testing, the SINet input is a test sample with historical values of F10.7 or F30
on the previous 30 days:
$d-30+1$, 
$d-30+2$, 
$\ldots$, 
$d-1$, 
$d$.
For the fixed prediction approach, the SINet output
consists of the predicted values of F10.7 or F30 for the next 60 days:
$d+1$, 
$d+2$,
$\ldots$,
$d+59$,
$d+60$.
For the rolling prediction approach, the SINet output is the predicted value of F10.7 or F30
on day $d+1$ (1 day in advance).
We then use the sliding window method described in Section \ref{sec:data} to iteratively predict
the values of F10.7 or F30
on day $d+2$ to day $d+60$.

Figure \ref{SINetarch}(b) presents the details of a TimesBlock. 
The input 1D time series data are first processed by a fast Fourier transform (FFT)
algorithm that converts the time domain to the frequency domain. 
The main purpose of the FFT algorithm is to identify and analyze periodic components within the data, specifically isolating the $k$ most significant periodic components,
where $k$ is a user-determined parameter.
(In the study presented here, $k$ is set to 3.)
 In this way, the model can focus on the important features to enhance learning effectiveness.
 After FFT, the $k$ most significant periods are reshaped into $k$ 2-dimensional (2D) tensors simultaneously, 
 which are then sent to a dual-inception model structure containing two inception blocks.
 The first inception block processes the reshaped data to extract hierarchical temporal features using multiscale 2D convolutional kernels. 
 These features are then refined by the second inception block, which further processes the data, reduces the dimensionality of the data, and prepares the data for reconstruction.
 After feature extraction by the inception blocks, the data are reshaped back into 1D time series, prepared for adaptive aggregation. 
 This dual-inception model structure enables TimesBlock to effectively capture both short- and long-term dependencies in the 1D time series data.

Separately, the output of the FFT algorithm is processed by a softmax function to calculate the softmax weights, which indicate the significance of each periodic component. 
The $k$ adaptive aggregation modules for the 
$k$ periodic components take as input the reshaped 1D time series data together with the softmax weights, 
and produce as output aggregated 1D time series data. 
Finally, the aggregated 1D time series data is combined with the original input of the TimesBlock through a residual connection. 
The residual connection prevents overfitting and improves generalization.

Figure \ref{SINetarch}(c) presents the details of the
dual-inception model structure, which is composed of two
inception blocks.
The first inception block consists of
six in-convolution blocks
(InConvBlock 1,
InConvBlock 2,
InConvBlock 3,
InConvBlock 4,
InConvBlock 5,
InConvBlock 6).
Each in-convolution block increases the dimension (i.e., the number of channels) in the block.
The second inception block consists of six out-convolution blocks
(OutConvBlock 1,
OutConvBlock 2,
OutConvBlock 3,
OutConvBlock 4,
OutConvBlock 5,
OutConvBlock 6).
Each out-convolution block decreases the dimension in the block.
The two inception blocks are connected by a
Gaussian error linear unit (Gelu) layer.
Table \ref{tab:dual-inception} presents the 
configuration details
of the dual-inception model structure for the medium-term daily prediction of the F10.7 and F30 solar indices 
(1-60 days in advance).
Using validation data for performance assessment, the best forecast result is achieved using historical values of the previous 30 days to predict the future values of F10.7
and F30 (1-60 days in advance).
Thus, the input (output, respectively) dimension of each 
in-convolution block
is represented by $30 \times C$ where $C = 32$ is the number of 
input channels 
($C = 64$ is the number of output channels, 
respectively) in the in-convolution block.
The input (output, respectively) dimension of each 
out-convolution block
is represented by $30 \times C$ where $C = 64$ is the number of 
input channels ($C = 32$ is the number of output channels, respectively) in the out-convolution block.

SINet is trained for 10 epochs, with a batch size of 32. We use the adaptive moment estimation (Adam) optimizer \citep{DBLP:journals/corr/KingmaB14} 
with a learning rate of 0.001, the mean square error (MSE) as the loss function, and a dropout rate of 0.3 to prevent overfitting. 
Table \ref{tab:hyper} summarizes the model training conditions
of SINet. 
The hyperparameter values displayed in Table \ref{tab:hyper} are obtained by using
the grid search capability from the Python machine learning library, {\tt scikit-learn} 
\citep{10.5555/1953048.2078195}.
The validation set used for tuning the hyperparameters 
is as described in Section \ref{sec:data}.
During training,
we input the training sample at each time point in the training set into the model to train the model.
When training is complete, the neuron weights in the model
are optimized.
During testing,
we input the test sample at each time point in the test set
into the trained model and calculate the model's prediction accuracy.
Training and test samples are also as described
in Section \ref{sec:data}.
The training, validation and testing processes were run
and completed
on an NVIDIA A100 GPU.

\begin{table}
\centering
\caption{Configuration Details of the Dual-Inception Model Structure in SINet}
\vspace*{-0.5cm}
\begin{tabular}{cccc}
\\ \hline
Layer              & Input Dimension       & Kernel Size  & Output Dimension       \\\hline
InConvBlock 1        & $30 \times 32$     & $1 \times 1$       & $ 30 \times 64$ \\
InConvBlock 2        & $30 \times 32$  &   $3 \times 3$       &  $ 30 \times 64$ \\
InConvBlock 3        & $30 \times 32$  & $5 \times 5$        & $ 30 \times 64$ \\
InConvBlock 4        & $30 \times 32$  & $7 \times 7$      &  $ 30 \times 64$\\
InConvBlock 5        & $30 \times 32$  & $9 \times 9$     &  $ 30 \times 64$ \\
InConvBlock 6        & $30 \times 32$ & $11 \times 11$     & $ 30 \times 64$\\
Activation (Gelu)  &  $ 30 \times 64$   & ---       &  $ 30 \times 64$ \\
OutConvBlock 1        &  $ 30 \times 64$& $1 \times 1$     &$30 \times 32$   \\
OutConvBlock 2       &  $ 30 \times 64$& $3 \times 3$    & $30 \times 32$   \\
OutConvBlock 3       &  $30 \times 64$ & $5 \times 5$  &$30 \times 32$\\
OutConvBlock 4       &  $ 30 \times 64$ & $7 \times 7$       & $30 \times 32$  \\
OutConvBlock 5      &  $ 30 \times 64$ &  $9 \times 9$       & $30 \times 32$  \\
OutConvBlock 6     &  $ 30\times 64$ & $11 \times 11$      & $30 \times 32$   
\\\hline
\end{tabular}
\label{tab:dual-inception}
\end{table}

\begin{table}
\centering
\caption{Model Training Conditions of SINet}
\vspace*{-0.5cm}
\begin{tabular}{cccccc}
\\ \hline
 Loss Function  & Optimizer & Learning Rate & Dropout Rate & Batch Size  & Epoch \\ \hline
 MSE            & Adam     & 0.001          & 0.3          & 32           & 10
\\ \hline 
\end{tabular}
\label{tab:hyper}
\end{table}

It should be pointed out that although both our SINet and the existing TimesNet \citep{DBLPWuHLZ0L23}
adopt TimesBlocks, they differ in several ways. 
First, TimesNet employs five most significant period components
in a TimesBlock, 
while SINet uses three most significant period components
to reduce model complexity.
Second, TimesNet uses SMAPE as the loss function
and a batch size of 16. 
In contrast, SINet uses MSE as the loss function
and a batch size of 32. 
In our case, the SMAPE loss function performs worse than MSE. 
Third, TimesNet is mainly designed for multivariate time series forecasting and employs a model dimension of 32.
In contrast, SINet is designed for univariate time series forecasting
specifically for F10.7 and F30 predictions, and adopts a model dimension of 64, which is more suitable for our work.
In contrast to 
other time series forecasting methods such as 
long short-term memory 
\citep[LSTM;][]{DBLP:journals/neco/HochreiterS97,
2022SoPh..297..157Z} 
and
temporal convolutional network 
\citep[TCN;][]{2024AnGeo..42...91W},
which process the input time series data directly,
SINet converts the input data from the time domain to
the frequency domain 
through a fast Fourier transform algorithm to enhance learning effectiveness.

\section{Results}
\label{sec:results}

\subsection{Evaluation Metrics}

We employ two different approaches with SINet:
fixed prediction, denoted SINet$_{\mbox{f}}$, and
rolling prediction, denoted SINet$_{\mbox{r}}$,
as explained in Section \ref{sec:data}.
We adopt three metrics to evaluate the performance of
the two approaches and related methods.
The three metrics are
the root mean square error (RMSE),
the mean absolute error (MAE), and the
mean absolute percentage error
\citep[MAPE;][]{DBLP:journals/ijon/MyttenaereGGR16}.

Let $N$ be the number of test samples.
For each test sample, we used the historical values of
F10.7 (F30, respectively) of the previous 30 days to
predict the values of F10.7 (F30, respectively)
in the future.
We focus on the $k$-day ahead prediction for
$k = 1, 27, 45, 60,$ respectively.
Let $\hat{y}_{i}$, $1 \leq i \leq N$, 
denote a predicted value, and let $y_{i}$ denote the corresponding
observed/true value. 
RMSE is calculated as:
\begin{equation}
    \mbox{RMSE} = \sqrt{{\frac{1}{N}}
    \sum_{i=1}^{N}{(\hat{y}_{i}-y_{i})^2}}.
\end{equation}
MAE is calculated as:
\begin{equation}
   \mbox{MAE} = {\frac{1}{N} 
   \sum_{i=1}^{N} |\hat{y}_{i}-y_{i}|}.
\end{equation}
MAPE is calculated as:
\begin{equation}
    \mbox{MAPE} = {\frac{1}{N} \sum_{i=1}^{N} 
    \frac{|\hat{y}_{i}-y_{i}|}{y_{i}} \times 100\%}.
\end{equation} 
The units of RMSE and MAE are sfu while the unit of MAPE is percentage. 
The smaller the RMSE, MAE and MAPE are, the better a method performs. 
As in \citet{jiang2025apjs}, we use MAPE as the primary metric.

\subsection{Experimental Results for F10.7}

\subsubsection{Performance Evaluation of SINet on F10.7 Prediction}
\label{sec:5foldF107}

We calculated the performance metric values of
SINet$_{\mbox{f}}$ and 
 SINet$_{\mbox{r}}$, and compared the metric values with those of five
 closely related forecasting methods
 for F10.7.
 The five methods include
autoregressive integrated moving average 
\citep[ARIMA;][]{box1970time}, 
long short-term memory 
\citep[LSTM;][]{DBLP:journals/neco/HochreiterS97,2022SoPh..297..157Z}, 
convolutional neural network 
\citep[CNN;][]{DBLP:journals/pieee/LeCunBBH98},
LSTM with multihead attention 
\citep[LSTM+;][]{DBLP:conf/nips/VaswaniSPUJGKP17,DBLP:journals/ascom/JerseM24}, 
and
temporal convolutional network 
\citep[TCN;][]{2024AnGeo..42...91W}.
 Among the five methods,
 ARIMA is a statistical learning method, while
 the other four are based on deep learning models.
 
 To increase the reliability of model evaluation,
 we conduct a 5-fold validation experiment.
To prevent data leakage and preserve the temporal order of the data, the training set and test set
in the 5-fold experiment are constructed as follows.
In fold 1, the training set and test set 
are as shown in Figure \ref{fig:data}.
In fold $i$, $2 \leq i \leq 5$, we remove the first $i-1$ months of data from
the training set in fold 1 to get a new training set
for fold $i$, and
remove the last $i-1$ months of data from the test set 
in fold 1
to get a new test set for fold $i$.
Thus, for fold $i$ and fold $j$, where $i \neq j$,
the training data and the test data in fold $i$
differ from the training data and the test data in fold $j$.
In each fold, we use the training data in that fold to train a model and
then use the test data in that fold to test the trained model,
to calculate the metric values in the fold.
Table \ref{tab:F107_Avg} presents the average and standard deviation of MAPE over the five folds
(more detailed results can be found in \ref{appendix}). 
Best metric values are highlighted in boldface.
It can be seen from Table \ref{tab:F107_Avg} that
SINet$_{\mbox{f}}$
performs better than the other methods, 
demonstrating its strong capability in capturing the temporal dynamics of solar flux variability.
TCN is generally the second best method.
For the 1-day ahead prediction,
SINet$_{\text{f}}$ and SINet$_{\text{r}}$
are identical methods.
For the larger forecast horizons,
SINet$_{\text{f}}$ outperforms SINet$_{\text{r}}$, indicating the advantage of the fixed prediction approach over
the rolling prediction approach. 
Results of the other metrics are similar and omitted.

\begin{table}[htbp]
\begin{small}
\centering
\caption{Average and Standard Deviation of MAPE (\%) Obtained by SINet 
and Related Methods on F10.7 Prediction in the 5-Fold Validation Experiment}
\begin{tabular}{lccccccc}
\hline
Forecast & SINet$_{\mbox{f}}$ & SINet$_{\mbox{r}}$ & ARIMA & LSTM & CNN & LSTM+ & TCN \\
\hline
$1^{st}$  Day
 & \textbf{2.3}$\pm0.1$ & \textbf{2.3}$\pm0.1$ & 2.4$\pm0.0$ & 2.8$\pm0.1$ & 3.0$\pm0.2$ & 2.5$\pm0.1$ & 2.4$\pm0.1$ \\
$27^{th}$ Day
 & \textbf{8.0}$\pm0.1$ & 8.8$\pm0.2$ & 9.1$\pm0.1$ & 8.9$\pm0.2$ & 8.5$\pm0.2$ & 8.7$\pm0.1$ & 8.6$\pm0.1$ \\
$45^{th}$ Day
 & \textbf{9.1}$\pm0.1$ & 9.3$\pm0.1$ & 11.3$\pm0.1$ & 11.0$\pm0.4$ & 9.6$\pm0.1$ & 9.6$\pm0.2$ & 9.4$\pm0.1$ \\
$60^{th}$ Day
 & \textbf{10.1}$\pm0.1$ & 10.9$\pm0.3$ & 11.6$\pm0.0$ & 11.1$\pm0.5$ & 10.7$\pm0.5$ & 10.4$\pm0.1$ & 10.4$\pm0.3$ \\
\hline
\end{tabular}
\label{tab:F107_Avg}
\end{small}
\end{table}

\subsubsection{Further Assessment of SINet and TCN on F10.7 Prediction}

\begin{figure}[ht!]
\centering
\includegraphics[width=0.9\linewidth]{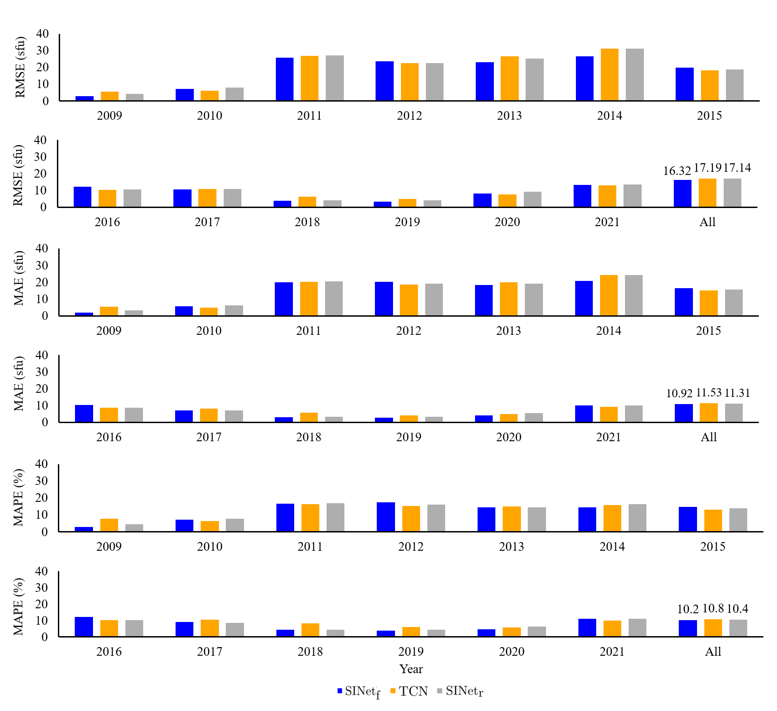}
\caption{Comparison of three forecasting methods:
SINet$_{\mbox{f}}$, TCN and SINet$_{\mbox{r}}$,
on the 60-day ahead prediction of F10.7
in the period between 2009 and 2021.
The figure shows the annual comparison results and the overall comparison results in this period.}
\label{fig:F107yearly}
\end{figure}

\begin{figure}[ht!]
\centering
\includegraphics[width=0.9\linewidth]{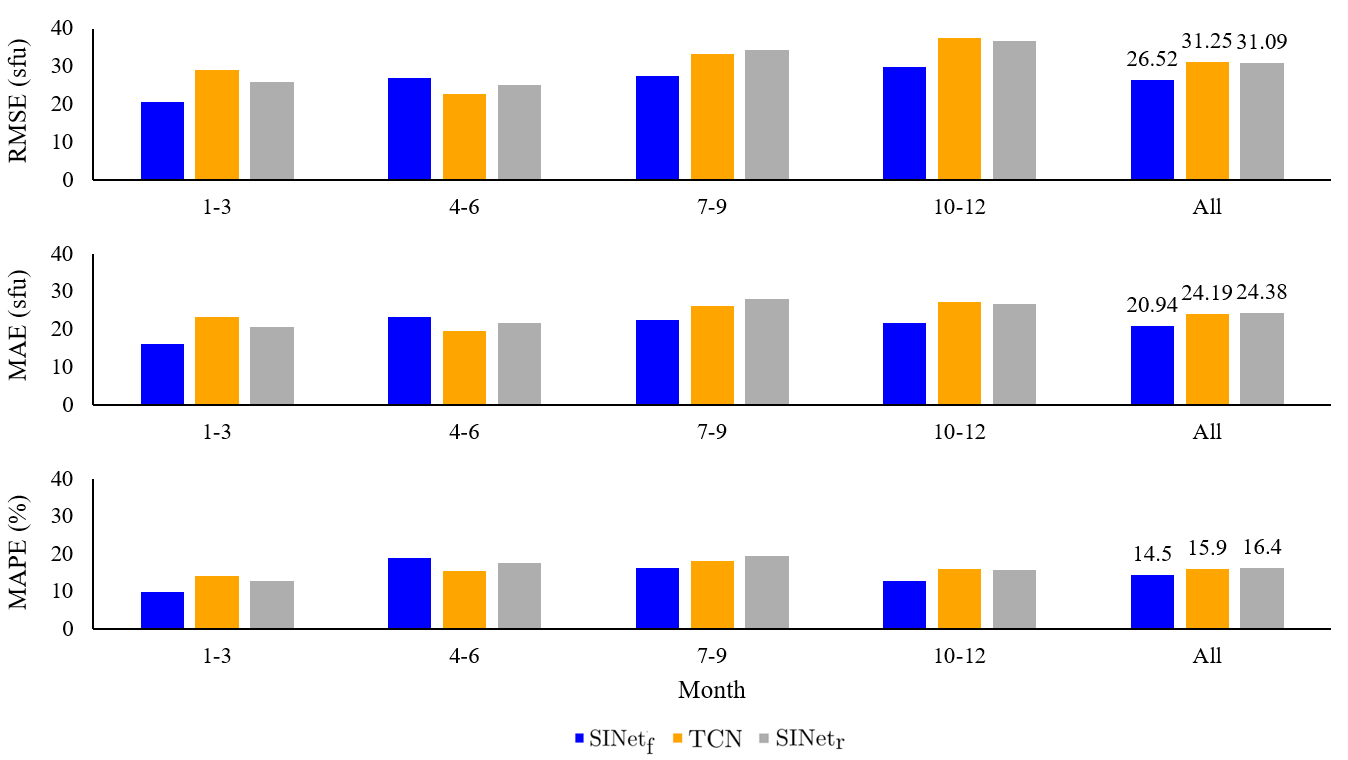}
\caption{Comparison of three forecasting methods:
SINet$_{\mbox{f}}$, TCN and SINet$_{\mbox{r}}$,
on the 60-day ahead prediction of F10.7
in the solar maximum (2014).
The figure shows 
the quarterly comparison results 
and the overall comparison results in 2014.}
\label{fig:F10solarmax}
\end{figure}

We further compared
the proposed approaches with the TCN method
on the 60-day ahead prediction of F10.7 
in the period between 2009 and 2021.
The training set and test set are as shown in 
Figure \ref{fig:data},
which are the same as those used in fold 1 
in Table \ref{tab:60thF107Comp}.
Figure \ref{fig:F107yearly} presents the annual comparison results and the overall comparison results of the three methods.
It can be seen in Figure \ref{fig:F107yearly} that the fixed prediction approach,
SINet$_{\mbox{f}}$, performs better than TCN and
the rolling prediction approach, SINet$_{\mbox{r}}$,
in the overall comparison results,
which are consistent with the results reported in Section \ref{sec:5foldF107}.
Specifically, SINet$_{\mbox{f}}$ achieves the lowest
RMSE of 16.32 sfu, MAE of 10.92 sfu, and MAPE of 10.2\%,
compared to RMSE of 17.19 sfu, MAE of 11.53 sfu, and MAPE of 10.8\%
(RMSE of 17.14 sfu, MAE of 11.31 sfu, and MAPE of 10.4\%, respectively)
obtained by TCN (SINet$_{\mbox{r}}$, respectively).
SINet$_{\mbox{f}}$ improves TCN by 5.06\% in RMSE,
5.29\% in MAE, and
5.56\% in MAPE.
As pointed out in Subsection 
\ref{sec:5foldF107},
the rolling prediction approach is worse than the fixed prediction approach.
This happens probably because, during the rolling, the historical values of the previous 30 days used to predict the 60-day ahead F10.7 value
are all synthetic F10.7 values rather than true F10.7 values. 

We note that in the solar maximum (i.e., 2014)
the three methods studied here are less accurate, and
our proposed SINet$_{\mbox{f}}$ is much better than
TCN in this period.
Figure \ref{fig:F10solarmax} provides a detailed comparison of
the three methods for daily predictions of F10.7
in 2014. 
It can be seen in Figure \ref{fig:F10solarmax} that
SINet$_{\mbox{f}}$ 
(TCN, SINet$_{\mbox{r}}$, respectively)
achieves an RMSE of 26.52 sfu
(31.25 sfu, 31.09 sfu, respectively),
an MAE of 20.94 sfu
(24.19 sfu, 24.38 sfu, respectively), and
an MAPE of 14.5\%
(15.9\%, 16.4\%, respectively) in 2014.
SINet$_{\mbox{f}}$ improves TCN by 15.14\% in RMSE,
13.44\% in MAE, and
8.81\% in MAPE in 2014.
There is a significant improvement of SINet$_{\mbox{f}}$ over TCN in the solar maximum.

\begin{figure}[ht!]
\centering
\includegraphics[width=0.8\linewidth]{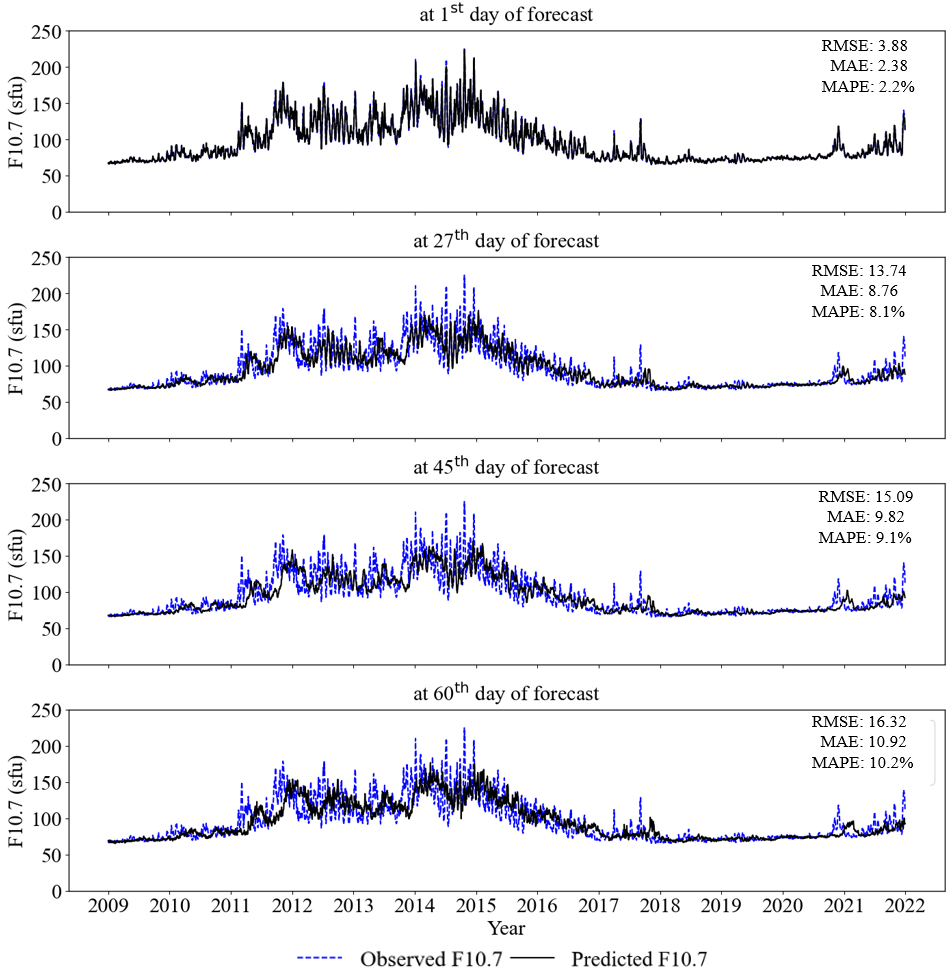}
\caption{Daily predictions of the F10.7 solar index 
made by our SINet$_{\mbox{f}}$ method
with four forecast horizons:
1, 27, 45, and 60 days, respectively,
in the period between 2009 and 2021.}
\label{fig:F107overview}
\end{figure}

Figure \ref{fig:F107overview} compares the observed/true values 
and the SINet$_{\mbox{f}}$-predicted F10.7 values
with four forecast horizons:
1, 27, 45 and 60 days, respectively,
in the period between 2009 and 2021.
Each dashed blue line represents the observed F10.7 values, 
while each solid black line represents the synthetic F10.7 values predicted by SINet$_{\mbox{f}}$.
In general, the SINet$_{\mbox{f}}$ method is
able to capture the trend of F10.7.
However, the larger the forecast horizon, the less accurate SINet$_{\mbox{f}}$ is. 
For 1-day ahead forecasts,
SINet$_{\mbox{f}}$ 
achieves an RMSE of 3.88 sfu,
an MAE of 2.38 sfu, and
an MAPE of 2.2\%.
For 60-day ahead forecasts, SINet$_{\mbox{f}}$ achieves an RMSE of 16.32 sfu, MAE of 10.92 sfu, and MAPE of 10.2\%.
The results of 60-day ahead forecasts 
are worse than those of 1-day ahead forecasts.

Recently, \citet{2024SoPh47Z} used a deep learning method, called Informer, to make daily predictions of F10.7
(1-27 days in advance).
When comparing SINet$_{\mbox{f}}$ with the Informer model
in the solar maximum (i.e., 2014),
the mean of RMSE, MAE, and MAPE, averaged over 1-27 forecast horizons,
for the Informer model is 23 sfu, 18 sfu, and 13\%, respectively.
The mean of RMSE, MAE, and MAPE, averaged over 1-27 forecast horizons,
for our SINet$_{\mbox{f}}$ method is
22 sfu, 17 sfu, and 11\%, respectively.
When comparing the mean values of annual average forecast errors with 
1-27 forecast horizons in the period between 2017 and 2021,
Informer achieves an RMSE of 6.285 sfu,
an MAE of 3.789 sfu,
and an MAPE of 4.54\%. 
SINet$_{\mbox{f}}$ achieves an RMSE of
5.881 sfu, 
an MAE of 3.843 sfu, and an MAPE of 4.53\%.
Although the Informer model and SINet$_{\mbox{f}}$ use different
prediction techniques with different training and test data,
these performance metric values indicate that our proposed approach is comparable to
the Informer model in making daily predictions of F10.7 
for 1-27 forecast horizons. 

\subsection{Experimental Results for F30}

\subsubsection{Performance Evaluation of SINet on F30 Prediction}
\label{sec:5foldF30}
Here, we calculated the performance metric values of 
SINet$_{\text{f}}$ and SINet$_{\text{r}}$
and compared the metric values with those of the five closely related
forecasting methods 
(ARIMA, LSTM, CNN, LSTM+, TCN)
for F30
using the 5-fold validation experiment
as described in
Section \ref{sec:5foldF107}.
Table \ref{tab:F30_Avg}
 presents the average and standard deviation of MAPE over the five folds
(more detailed results can be found in \ref{appendix}).
Again, 
Table \ref{tab:F30_Avg}
shows that
SINet$_{\mbox{f}}$
is the best method while 
TCN is generally the second best method in terms of MAPE.
Results of the other metrics are similar and omitted.

\begin{table}[htbp]
\begin{small}
\centering
\caption{Average and Standard Deviation of MAPE (\%) Obtained by SINet and Related Methods 
on F30 Prediction
in the 5-Fold Validation Experiment}
\begin{tabular}{lccccccc}
\hline
Forecast & SINet$_{\mbox{f}}$ & SINet$_{\mbox{r}}$ & ARIMA & LSTM & CNN & LSTM+ & TCN \\
\hline
$1^{st}$ Day
 & \textbf{2.0}$\pm0.0$ & \textbf{2.0}$\pm0.0$ & 2.1$\pm0.0$ & 2.6$\pm0.2$ & 3.4$\pm0.2$ & 2.4$\pm0.2$ & 2.4$\pm0.2$ \\
$27^{th}$ Day
 & \textbf{7.1}$\pm0.1$ & 7.9$\pm0.1$ & 7.6$\pm0.1$ & 8.5$\pm0.1$ & 8.2$\pm0.1$ & 7.6$\pm0.1$ & 7.5$\pm0.2$ \\
$45^{th}$ Day
 & \textbf{8.7}$\pm0.1$ & 8.9$\pm0.2$ & 10.5$\pm0.1$ & 10.6$\pm0.1$ & 9.0$\pm0.3$ & 9.9$\pm0.2$ & 9.2$\pm0.3$ \\
$60^{th}$ Day
 & \textbf{9.5}$\pm0.1$ & 10.2$\pm0.3$ & 10.7$\pm0.0$ & 10.3$\pm0.1$ & 10.2$\pm0.1$ & 9.6$\pm0.1$ & 9.6$\pm0.1$ \\
\hline
\end{tabular}
\label{tab:F30_Avg}
\end{small}
\end{table}

\subsubsection{Further Assessment of SINet and TCN on F30 Prediction}
\begin{figure}[t!]
\centering
\includegraphics[width=0.9\linewidth]{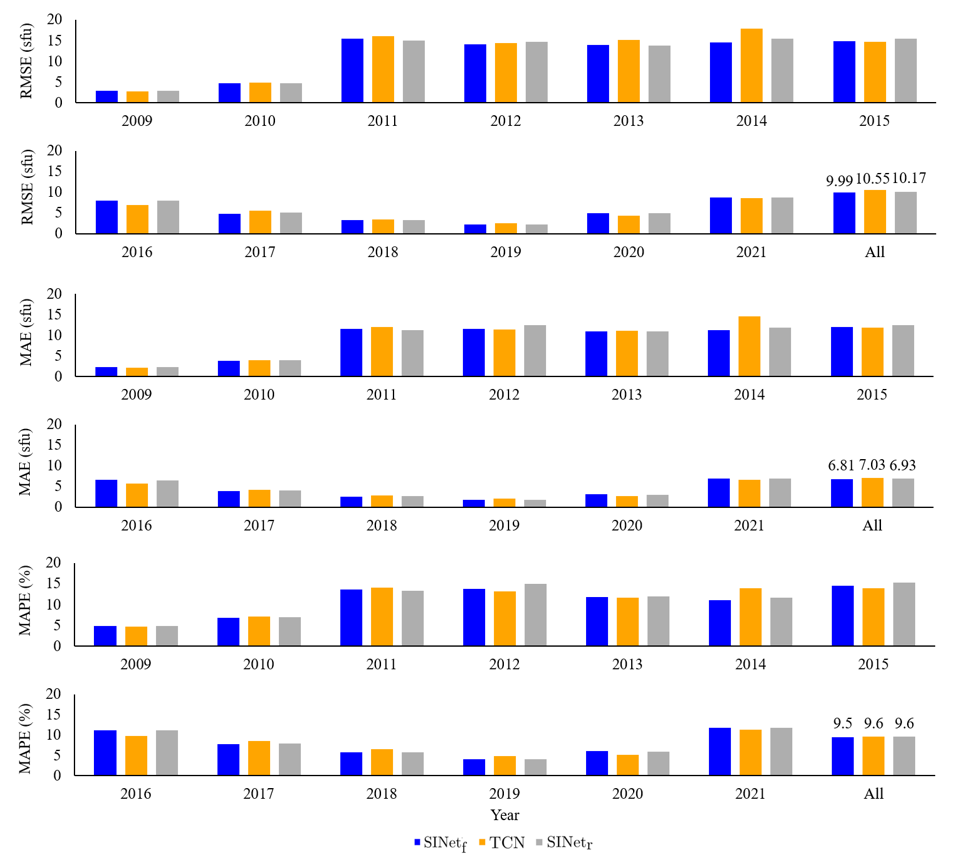}
\caption{Comparison of three forecasting methods:
SINet$_{\mbox{f}}$, TCN and SINet$_{\mbox{r}}$,
on the 60-day ahead prediction of F30
in the period between 2009 and 2021.
The figure shows the annual comparison results and the overall comparison results in this period.}
\vspace*{-0.5cm}
\label{fig:F30yearly}
\end{figure}

We further compared
the proposed approaches with the TCN method
on the 60-day ahead prediction of F30 
in the period between 2009 and 2021.
The training set and test set are as shown in 
Figure \ref{fig:data},
which are the same as those used in fold 1 
in Table \ref{tab:60thF30Comp}.
Figure \ref{fig:F30yearly} presents the annual comparison results and the overall comparison results of the three methods.
As in the F10.7 case,
SINet$_{\mbox{f}}$ performs the best
in the overall comparison results on F30 prediction.
SINet$_{\mbox{f}}$ achieves an
RMSE of 9.99 sfu, MAE of 6.81 sfu, and MAPE of 9.5\%,
compared to the RMSE of 10.55 sfu, MAE of 7.03 sfu, and MAPE of 9.6\%
(RMSE of 10.17 sfu, MAE of 6.93 sfu, and MAPE of 9.6\%, respectively)
obtained by
TCN (SINet$_{\mbox{r}}$, respectively).
SINet$_{\mbox{f}}$ improves TCN by 5.31\% in RMSE,
3.13\% in MAE, and
1.04\% in MAPE.
Comparing Figure \ref{fig:F30yearly} with 
Figure \ref{fig:F107yearly},
we see that SINet$_{\mbox{f}}$ achieves
higher accuracy
in F30 than in F10.7.
This happens because,
in the test set,
the range of the F30 values, which are
38.9 sfu to 199.1 sfu, is much smaller
than the range of the F10.7 values,
which are 65.7 sfu to 371.1 sfu.
A smaller range with lower variability causes smaller prediction errors.

\begin{figure}
\centering
\includegraphics[width=0.9\linewidth]{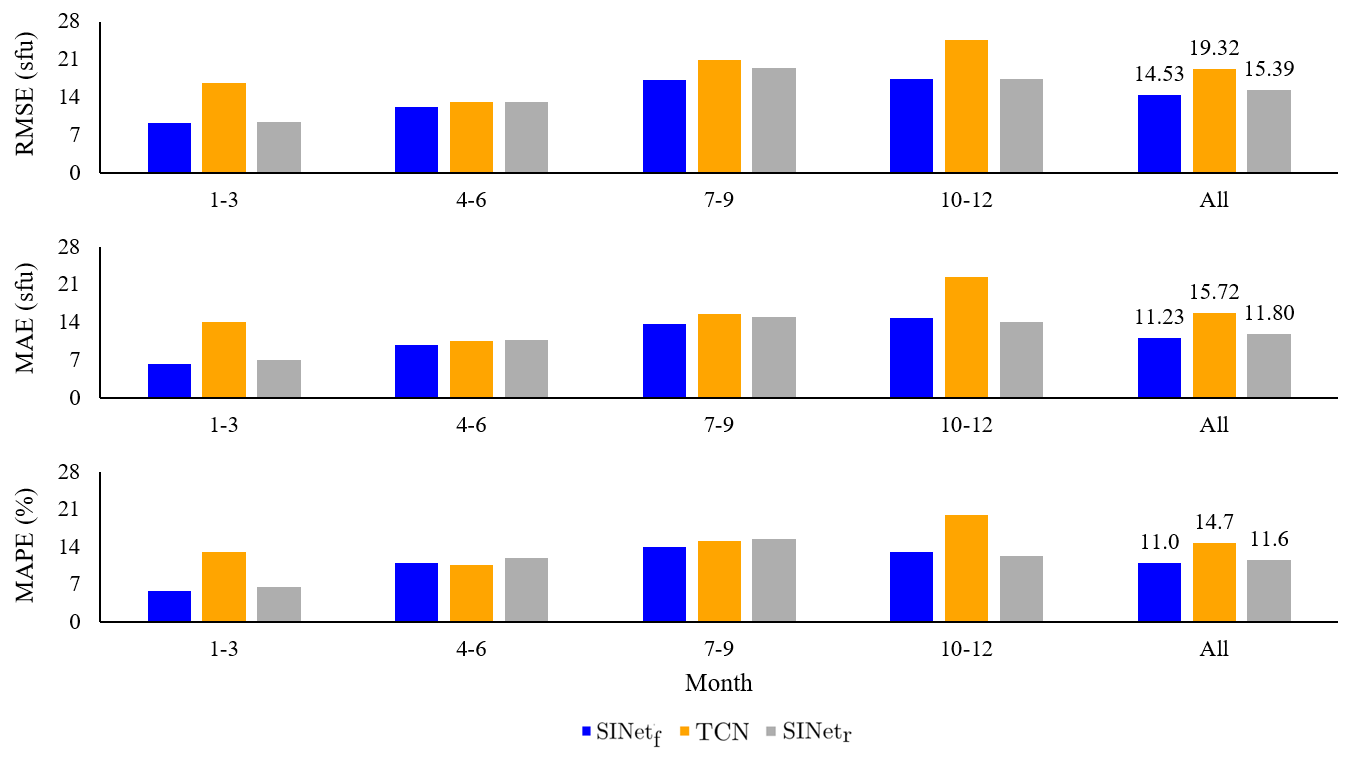}
\caption{Comparison of three forecasting methods:
SINet$_{\mbox{f}}$, TCN and SINet$_{\mbox{r}}$,
on the 60-day ahead prediction of F30
in the solar maximum (2014).
The figure shows 
the quarterly comparison results 
and the overall comparison results in 2014.}
\vspace*{-0.5cm}
\label{fig:F30solarmax}
\end{figure}

Again, we note that in the solar maximum (i.e., 2014)
the three methods studied here are less accurate, and
our proposed SINet$_{\mbox{f}}$ is much better than
TCN in this period.
Figure \ref{fig:F30solarmax} provides a detailed comparison of
the three methods for daily predictions of F30
in 2014.
It can be seen in Figure \ref{fig:F30solarmax} that
SINet$_{\mbox{f}}$ 
(TCN, SINet$_{\mbox{r}}$, respectively)
achieves an RMSE of 14.53 sfu
(19.32 sfu, 15.39 sfu, respectively),
an MAE of 11.23 sfu
(15.72 sfu, 11.80 sfu, respectively),
an MAPE of 11.0\% (14.7\%, 11.6\%, respectively).
SINet$_{\mbox{f}}$ improves TCN by 
24.79\% in RMSE,
28.56\% in MAE, and
25.17\% in MAPE.
Again, there is a significant improvement of SINet$_{\mbox{f}}$ over TCN
in the solar maximum.

\begin{figure}
\centering
\includegraphics[width=0.8\linewidth]{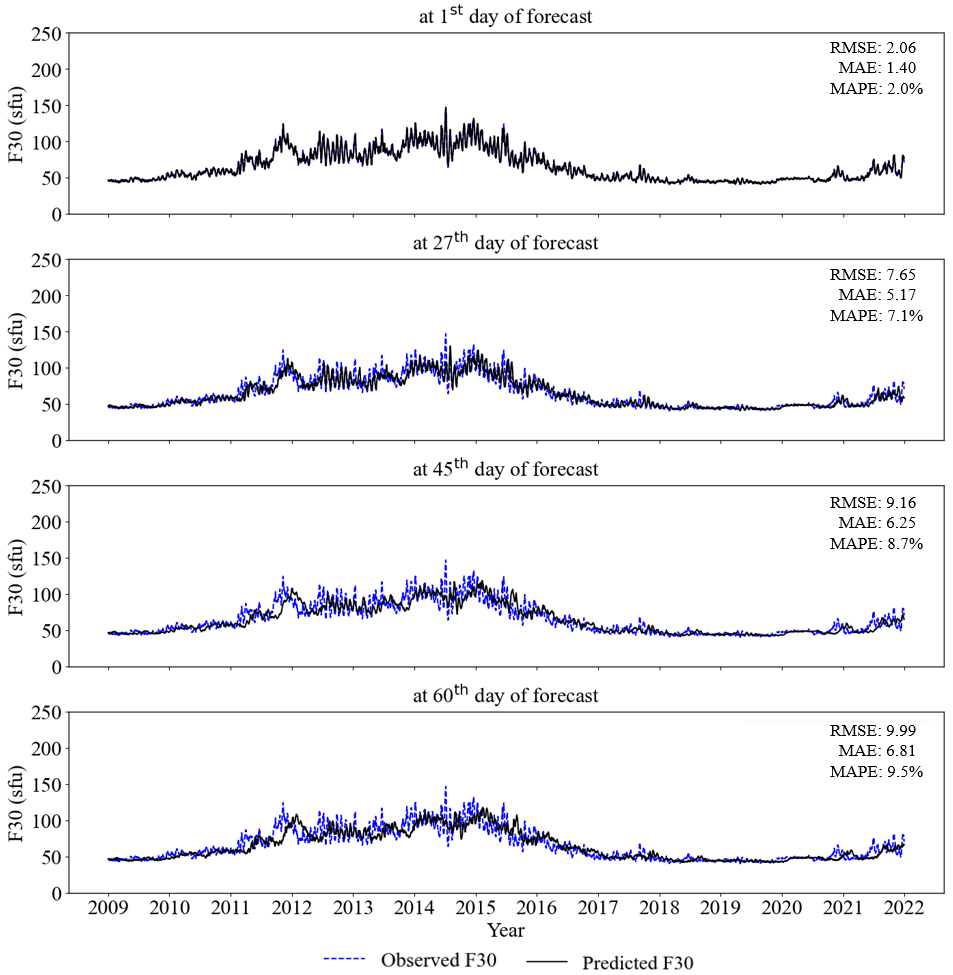}
\caption{Daily predictions of the F30 solar index 
made by our SINet$_{\mbox{f}}$ method
with four forecast horizons:
1, 27, 45, and 60 days, respectively,
in the period between 2009 and 2021.}
\label{fig:F30overview}
\end{figure}

Figure \ref{fig:F30overview} compares the observed/true values 
and the SINet$_{\mbox{f}}$-predicted F30 values
with four forecast horizons:
1, 27, 45, and 60 days, respectively,
in the period between 2009 and 2021.
Each dashed blue line represents the observed F30 values,
while each solid black line represents the synthetic F30 values
predicted by SINet$_{\mbox{f}}$.
Like the F10.7 case, the SINet$_{\mbox{f}}$ method is
able to capture the trend of F30.
However, the larger the forecast horizon, the less accurate SINet$_{\mbox{f}}$ is.
For 1-day ahead forecasts,
SINet$_{\mbox{f}}$ 
achieves an RMSE of 2.05 sfu,
an MAE of 1.40 sfu, and
an MAPE of 2.0\%.
For 60-day ahead forecasts, SINet$_{\mbox{f}}$ achieves an RMSE of 9.99 sfu, MAE of 6.81 sfu, and MAPE of 9.5\%.
Comparing Figure \ref{fig:F30overview} with Figure \ref{fig:F107overview},
we see that the predicted values of F30 are closer to the observed/true values
than those of F10.7.

\subsection{Daily Predictions During the Period of Active Region NOAA 12673}

\begin{figure}[t!]
\centering
\includegraphics[width=0.9\linewidth]{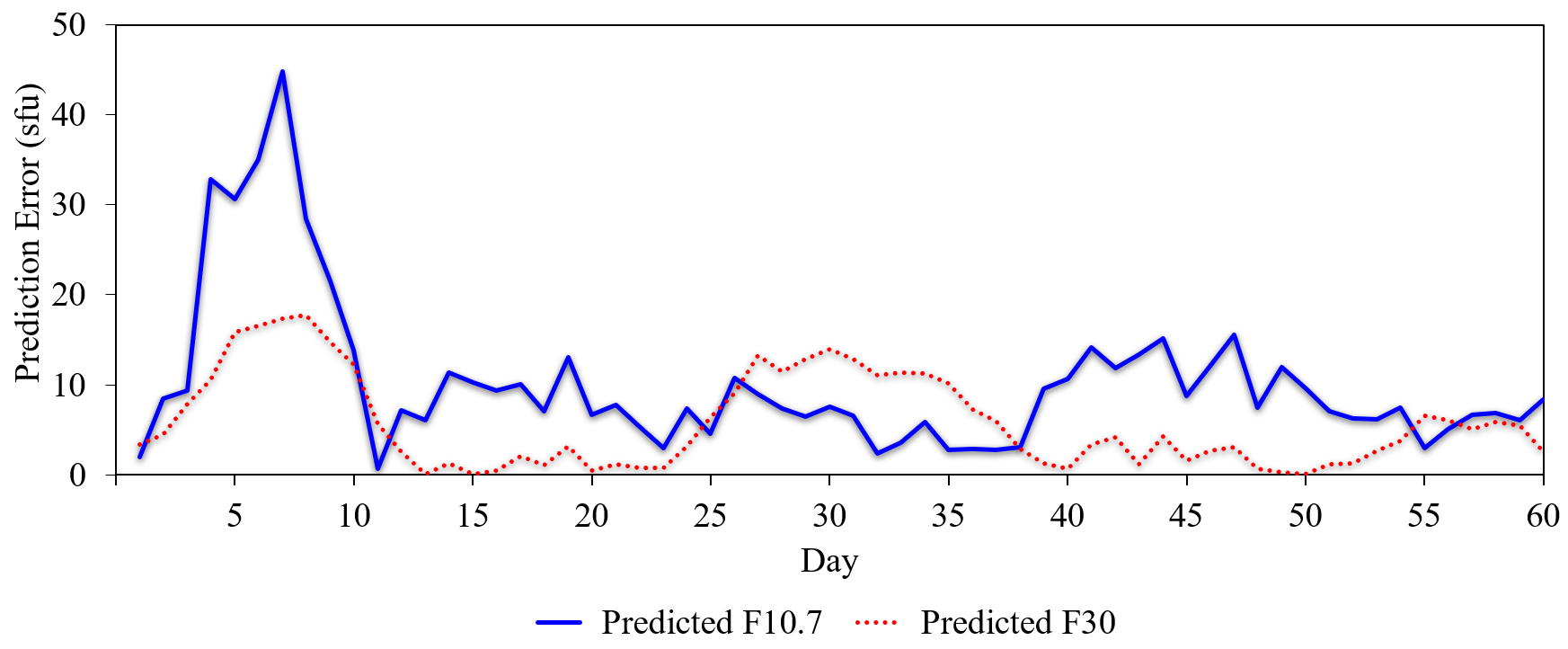}
\caption{Prediction errors 
made by SINet$_{\mbox{f}}$
for the F10.7 and F30 solar indices
in the period between September 1, 2017 and October 30, 2017
(that is, between day 1 and day 60 on the X-axis).}
\vspace*{-0.4cm}
\label{fig:2017_60day}
\end{figure}

Active Region (AR) NOAA 12673
was most active in September 2017, specifically between September 4, 2017 and September 10, 2017.
It was a super-flaring active region that produced numerous solar flares, including a very large X9.3 flare. This region showed exceptionally fast magnetic flux emergence and was characterized by a complex interplay of magnetic field and dynamics. 
One wonders how the SINet$_{\mbox{f}}$ method performs during the very dynamic AR 12673 period.

Our trained SINet$_{\mbox{f}}$ model was fed
with solar index values in the period between
August 2, 2017 and August 31, 2017
and made daily predictions of F10.7 and F30 in
the period between September 1, 2017 and October 30, 2017
(1-60 days in advance).
Figure \ref{fig:2017_60day} shows the results,
in which the prediction error on a day
is defined as the absolute value of the
difference between the true/observed value and
the SINet$_{\mbox{f}}$-predicted value on that day.
The solid blue line represents the prediction errors for F10.7 and the dotted red line represents the prediction errors for F30. 
It can be seen in Figure \ref{fig:2017_60day} that
the prediction errors are higher
in the period
between September 4, 2017 and September 10, 2017
(that is, between day 4 and day 10 on the X-axis)
than in the other periods.
This happens probably because there are relatively few
training samples related to such a fast magnetic flux emergence
as in
the dynamic AR 12673 period, and hence
its pattern is not captured by the
SINet$_{\mbox{f}}$ model during training.
Consequently, during testing/prediction, the
pattern can not be recognized, and hence the model
does not perform well in this period.

This case study shows that our 
SINet$_{\mbox{f}}$
model
suffers when encountering the emergence of new, very active ARs where
the AR emergence is much more a random process than the AR decay process.
This is a limitation of all solar activity forecasting models.
Overcoming this limitation can be challenging.
Literature suggests several ways to 
alleviate this challenging problem,
which include removing solar transient events such as 
very large flares and coronal mass
ejections from the dataset 
\citep{jiang2025apjs}, or
using oversampling \citep{2025ApJ...981...37Z} and
data augmentation techniques \citep{2025ApJS..280...52L}
to increase model's knowledge of the most active ARs.
Another possible way is to create a separate machine learning model,
such as an anomaly prediction model
\citep{DBLP:journals/access/FahimS19},
dedicated to solar index forecasting during solar transient events. 

\section{Discussion}
\label{sec:discussion}

The input length of our
SINet$_{\mbox{f}}$ model
is 30 days,
which is approximately a solar rotation period
(see Figure \ref{fig:fixrolling}).
It is known that solar variability can be characterized with a 7-month long impulse function; specifically, after an AR emerges, its impact on solar irradiance decays over 7 solar rotations
\citep[e.g.,][]{2018ApJ...853..197D,2005GeoRL..3214109P}.
It is worthwhile to consider using more than 1-month data for 
the 
model input so that our model might do better at capturing
the known solar physical effect of AR impacts on solar irradiance variability.
We conducted additional experiments to evaluate the impact of the input length on model performance.
It was found that using input data of more months (e.g. 2-month, 3-month, 4-month, etc.)
produced similar results, though model training time increased substantially.
Information in one solar rotation period suffices for the model
to make decisions on solar irradiance predictions. 
Similarly,
we selected the 3 most significant periodic components
when using the fast
Fourier transform algorithm to convert the input data from the time domain to the frequency domain
(see Figure \ref{SINetarch}).
Increasing the number of components
produced similar results while increasing model
training time.
On the other hand, selecting too few components
(e.g., 1 or 2 components)
led to worse performance.

Figure \ref{fig:F107yearly}
and Figure \ref{fig:F30yearly} show that
our SINet$_{\mbox{f}}$ model has
the best metrics averaged over the 13 years
from 2009 to 2021, but is worse 
than the state-of-the-art TCN model 
in some individual years (e.g., 2016). This happens probably because TCN is designed to capture temporal patterns in time series. In 2016, TCN can better capture such patterns in solar index time series with less variability when solar activity is weaker. 
However, in the solar maximum (2014) or in
the solar minimum (2019) during which 
such temporal patterns do not exist,
SINet$_{\mbox{f}}$ shows better inference 
capability and 
achieves better performance than TCN.

Figure \ref{fig:F107overview} and
Figure \ref{fig:F30overview}
show that SINet$_{\mbox{f}}$ has an autocorrelation basis;
specifically, past solar rotation is largely duplicated into the forecast.
We investigated this autocorrelation effect
by considering
predictions with a forecast horizon of 1 day since
these predictions are most accurate.
Mathematically, the autocorrelation 
at lag $k$, denoted $\hat{\rho}_k$, is defined as
\begin{equation}
\hat{\rho}_k =
\frac{\sum_{t = k + 1}^{N}
\left( \hat{y}_t - \overline{\hat{y}} \right)
\left( \hat{y}_{t - k} - \overline{\hat{y}} \right)}
{\sum_{t = 1}^{N}
\left( \hat{y}_t - \overline{\hat{y}} \right)^2},
\quad
k = 1, 27, 45, 60,
\end{equation}
where $\hat{y}_t$ denotes the 
predicted value at time point $t$,
$\overline{\hat{y}}$ is the mean of the predicted series, 
and $N$ is the total number of time points.
The statistic $\hat{\rho}_k$ measures the autocorrelation between the
predicted value $\hat{y}_t$ and its $k$-day lagged version 
$\hat{y}_{t-k}$.

\begin{figure}[t!]
\centering
\includegraphics[width=0.9\linewidth]{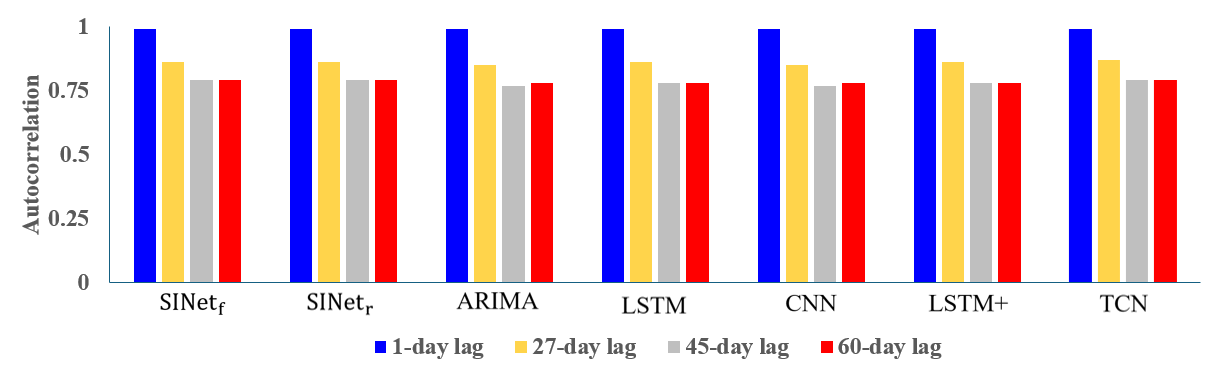}
\caption{Comparison of autocorrelation
effects among seven forecasting methods on
the 1-day ahead prediction of F10.7 in the
period between 2009 and 2021.}
\label{fig:F10autocorr}
\end{figure}

\begin{figure}[t!]
\centering
\includegraphics[width=0.9\linewidth]{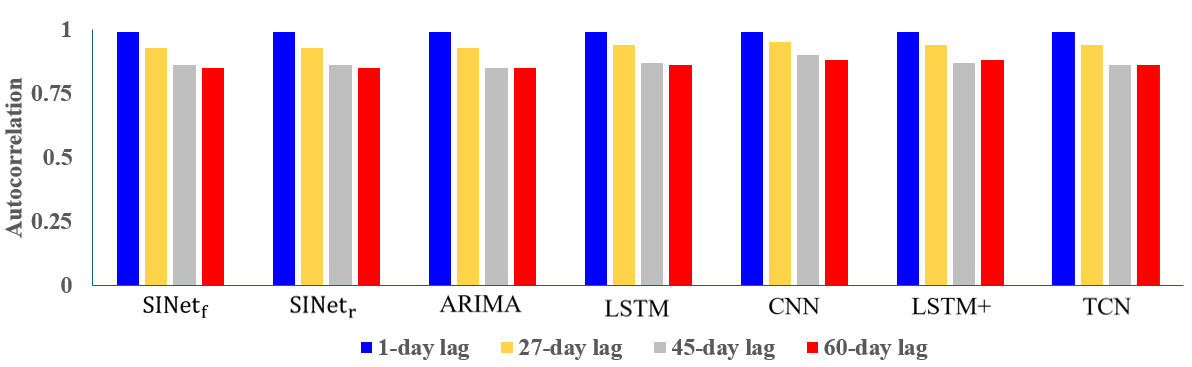}
\caption{Comparison of autocorrelation
effects among seven forecasting methods on
the 1-day ahead prediction of F30 in the
period between 2009 and 2021.
}
\label{fig:F30autocorr}
\end{figure}

Figure \ref{fig:F10autocorr} 
(Figure \ref{fig:F30autocorr}, respectively)
presents results for
F10.7
(F30, respectively).
It can be seen from the figures
that there is a clear autocorrelation effect
at lag 1 for all the seven methods
studied here.
This result is understandable given that the one-day
ahead prediction is highly accurate,
as shown in Sections \ref{sec:5foldF107} 
and \ref{sec:5foldF30},
and there is not much change in solar irradiance
within one day.
In contrast, when $k$ is larger, 
the autocorrelation effect is less evident,
suggesting that the predicted values are less autocorrelated over a longer period,
as solar activity may vary 
in the period.

\section{Conclusion}
\label{sec:conclusion}

In this study, we presented a novel deep learning model (SINet) for
making medium-term daily predictions of the F10.7 and F30 solar indices (1-60 days in advance).
SINet is an enhancement of TimesNet  \citep{DBLPWuHLZ0L23}, 
specifically designed to capture short- and long-range dependencies in
the time series of F10.7 and F30 values.
The model takes advantage of the frequency domain decomposition via
a fast Fourier transform (FFT) algorithm with a dual-inception model
structure, enabling effective extraction of dominant periodic components
in the time series data. 

We adopted fixed and rolling prediction approaches,
denoted
SINet$_{\mbox{f}}$
and
SINet$_{\mbox{r}}$,
respectively,
and compared them with
five closely related statistical and deep learning methods
(ARIMA, LSTM, CNN, LSTM+, TCN).
Extensive experiments show that SINet$_{\mbox{f}}$ performs the best
while TCN is generally
the second best method. 
When predicting the F10.7 solar index (60 days in advance), 
TCN achieves an RMSE of 17.19 sfu, MAE of 11.53 sfu, and MAPE of 10.8\%,
in the period between 2009 and 2021.
In contrast,
SINet$_{\mbox{f}}$ achieves an RMSE of 16.32 sfu, MAE of 10.92 sfu, and MAPE of 10.2\%, in the same period.
When predicting the F30 solar index (60 days in advance),
TCN achieves an RMSE of 10.55 sfu, MAE of 7.03 sfu, and MAPE of 9.6\%,
in this period.
In contrast, 
SINet$_{\mbox{f}}$ achieves an RMSE 9.99 sfu, MAE of 6.81 sfu, and MAPE of 9.5\%,
in the period.
In the solar maximum (2014),
TCN achieves an RMSE of 31.25 sfu, MAE of 24.19 sfu, and MAPE of 15.9\%
for F10.7 and
an RMSE of 19.32 sfu, MAE of 15.72 sfu, and MAPE of 14.7\%
for F30.
In contrast, 
SINet$_{\mbox{f}}$ achieves an RMSE of 26.52 sfu, MAE of 20.94 sfu, and MAPE of 14.5\%
for F10.7 and
an RMSE of 14.53 sfu, MAE of 11.23 sfu, and MAPE of 11.0\%
for F30.
SINet$_{\mbox{f}}$ improves the state-of-the-art TCN 
method significantly during the solar maximum
for 60-day ahead forecasts.
Based on these results, we conclude that
SINet$_{\mbox{f}}$ is a feasible model to make
medium-term daily predictions of the F10.7 and F30 solar indices.

We have implemented SINet$_{\mbox{f}}$ into an
operational 
F10.7
forecasting system.
Visit\\
\url{https://nature.njit.edu/solardb/},
click the ``Tools'' tab and select "F10.7 Forecasting" in the drop-down menu to access the 
SINet$_{\mbox{f}}$ system.
The National Oceanic and Atmospheric Administration (NOAA)
Space Weather Prediction Center (SWPC)
also provides an operational 
F10.7
forecasting system, 
which produces 45-day forecasts of F10.7 at
\url{https://www.swpc.noaa.gov/products/usaf-45-day-ap-and-f107cm-flux-forecast}
\citep{2017SpWea..15.1039W}.
We compared our SINet$_{\mbox{f}}$ system with
the NOAA/SWPC forecasting system
in the period between
March 29, 2025 and June 24, 2025,
during which observed F10.7 values
and archived F10.7 predictions at NOAA/SWPC
are available.
It was found that,
on the operational F10.7 predictions,
the MAPE of the SINet$_{\mbox{f}}$ system 
is 13.8\%, which is slightly better than the MAPE of 14.0\% achieved by the NOAA/SWPC forecasting system,
suggesting that
our SINet$_{\mbox{f}}$ system
can complement the existing \\
NOAA/SWPC forecasting system.

\acknowledgments
We are grateful to the anonymous referees for their valuable comments and constructive suggestions
that have helped improve the manuscript
significantly.
We also thank members of the Institute for Space Weather Sciences
for fruitful discussions.
The SINet model was implemented in PyTorch.
This work was supported in part by
NSF grants
AGS-2149748, 
AGS-2228996, 
AGS-2300341,
OAC-2320147, 
RISE-2425602,
OAC-2504860, and NASA grants 80NSSC24K0548,
80NSSC24K0843, 
and 
80NSSC24M0174.\\

\vspace*{+0.65cm}

\noindent
\textbf{Conflict of Interest}\\
The authors declare no conflicts of interest relevant to this study.

\appendix
\section{Detailed 5-Fold Validation Results for F10.7 and F30}
\label{appendix}

Table \ref{tab:1stF107Comp}
(Table \ref{tab:27thF107Comp},
Table \ref{tab:45thF107Comp}, and
Table \ref{tab:60thF107Comp}, respectively)
presents 
MAPE results
of an F10.7 forecasting method
obtained in each fold of the 5-fold validation experiment
at the $1^{st}$ day of forecast
(the $27^{th}$ day of forecast,
the $45^{th}$ day of forecast, and
the $60^{th}$ day of forecast, respectively).
Table~\ref{tab:1stF30Comp}
(Table \ref{tab:27thF30Comp},
Table \ref{tab:45thF30Comp}, and
Table \ref{tab:60thF30Comp}, respectively)
presents MAPE results
of an F30 forecasting method
obtained in each fold of the 5-fold validation experiment
at the $1^{st}$ day of forecast
(the $27^{th}$ day of forecast,
the $45^{th}$ day of forecast, and
the $60^{th}$ day of forecast, respectively).
Each MAPE value of a forecasting method is associated with 
a $p$-value obtained by applying
the Wilcoxon signed-rank test
\citep{Wilcoxon1945}
to the forecasting method and 
SINet$_{\mbox{f}}$, where
a $p$-value less than or equal to 0.05 indicates a statistically significant result.
For the 1-day ahead prediction,
SINet$_{\text{f}}$ and SINet$_{\text{r}}$
are identical methods.

\begin{table}[htbp]
\begin{small}
\centering
\caption{MAPE (\%) of SINet and Related Methods on F10.7 Prediction (at $1^{st}$ Day of Forecast)}
\begin{tabular}{lccccccc}
\hline
 & SINet$_{\mbox{f}}$ & SINet$_{\mbox{r}}$ & ARIMA & LSTM & CNN & LSTM+ & TCN \\ 
\hline
Fold 1 & \textbf{2.2} & \textbf{2.2} & 2.4 & 2.8 & 2.4 & 2.4 & 2.3 \\
$p$-value &  & --- & 2.9E-2 & 8.3E-70 & 1.3E-31 & 7.1E-26 & 1.7E-39 \\
Fold 2 & \textbf{2.3} & \textbf{2.3} & 2.4 & 3.1 & 2.4 & 2.6 & 2.4 \\
$p$-value &  & --- & 9.5E-2 & 1.5E-83 & 9.6E-10 & 1.7E-62 & 5.9E-15 \\
Fold 3 & \textbf{2.3} & \textbf{2.3} & 2.4 & 2.6 & 2.8 & 2.6 & 2.4 \\
$p$-value &  & --- & 1.3E-1 & 2.3E-50 & 1.9E-90 & 7.8E-47 & 3.9E-16 \\
Fold 4 & \textbf{2.2} & \textbf{2.2} & 2.4 & 3.1 & 2.8 & 2.4 & 2.4 \\
$p$-value &  & --- & 8.1E-2 & 9.8E-51 & 1.1E-93 & 1.1E-13 & 1.7E-31 \\
Fold 5 & \textbf{2.2} & \textbf{2.2} & 2.4 & 3.1 & 2.8 & 2.4 & 2.4 \\
$p$-value &  & --- & 1.4E-2 & 7.1E-112 & 3.7E-139 & 5.5E-27 & 3.1E-46 \\
\hline
\end{tabular}
\label{tab:1stF107Comp}
\end{small}
\end{table}

\begin{table}[htbp]
\begin{small}
\centering
\caption{MAPE (\%) of SINet and Related Methods on F10.7 Prediction (at $27^{th}$ Day of Forecast)}
\begin{tabular}{lccccccc}
\hline
 & SINet$_{\mbox{f}}$ & SINet$_{\mbox{r}}$ & ARIMA & LSTM & CNN & LSTM+ & TCN \\
\hline
Fold 1 & \textbf{8.1} & 8.7 & 9.2 & 8.7 & 8.7 & 8.8 & 8.7 \\
$p$-value &  & 4.6E-25 & 2.8E-10 & 2.7E-15 & 9.8E-30 & 1.8E-51 & 7.7E-105 \\
Fold 2 & \textbf{8.0} & 8.8 & 9.1 & 8.8 & 8.5 & 8.7 & 8.7 \\
$p$-value &  & 1.6E-23 & 1.6E-11 & 3.8E-19 & 2.4E-44 & 7.7E-27 & 3.1E-21 \\
Fold 3 & \textbf{8.1} & 9.0 & 9.1 & 8.9 & 8.4 & 8.8 & 8.5 \\
$p$-value &  & 2.6E-18 & 1.5E-9 & 1.8E-18 & 3.5E-24 & 1.1E-69 & 4.9E-55 \\
Fold 4 & \textbf{8.0} & 8.6 & 9.1 & 9.2 & 8.3 & 8.7 & 8.6 \\
$p$-value &  & 5.2E-17 & 2.7E-10 & 7.9E-31 & 2.7E-20 & 6.7E-24 & 6.7E-31 \\
Fold 5 & \textbf{8.0} & 9.0 & 9.1 & 9.1 & 8.6 & 8.7 & 8.5 \\
$p$-value &  & 3.0E-28 & 8.3E-12 & 5.0E-63 & 6.3E-25 & 1.4E-20 & 1.1E-54 \\
\hline
\end{tabular}
\label{tab:27thF107Comp}
\end{small}
\end{table}

\begin{table}[htbp]
\begin{small}
\centering
\caption{MAPE (\%) of SINet and Related Methods on F10.7 Prediction (at $45^{th}$ Day of Forecast)}
\begin{tabular}{lccccccc}
\hline
 & SINet$_{\mbox{f}}$ & SINet$_{\mbox{r}}$ & ARIMA & LSTM & CNN & LSTM+ & TCN \\
\hline
Fold 1 & \textbf{9.1} & 9.2 & 11.4 & 11.3 & 9.7 & 9.8 & 9.4 \\
$p$-value &  & 1.4E-1 & 5.5E-16 & 3.7E-207 & 1.5E-10 & 3.7E-70 & 8.1E-45 \\
Fold 2 & \textbf{9.1} & 9.3 & 11.3 & 10.7 & 9.7 & 9.3 & 9.3 \\
$p$-value &  & 8.3E-1 & 2.2E-18 & 2.4E-142 & 1.7E-7 & 1.1E-83 & 1.7E-36 \\
Fold 3 & \textbf{9.2} & 9.3 & 11.3 & 11.3 & 9.7 & 9.5 & 9.4 \\
$p$-value &  & 4.7E-3 & 2.6E-16 & 2.4E-175 & 1.1E-17 & 1.3E-154 & 3.7E-50 \\
Fold 4 & \textbf{9.2} & 9.3 & 11.3 & 11.3 & 9.6 & 9.5 & 9.5 \\
$p$-value &  & 6.5E-1 & 1.1E-16 & 1.8E-183 & 4.4E-6 & 1.9E-57 & 1.5E-11 \\
Fold 5 & \textbf{9.2} & 9.4 & 11.3 & 10.5 & 9.6 & 9.7 & 9.3 \\
$p$-value &  & 3.1E-11 & 4.8E-16 & 2.6E-263 & 2.9E-7 & 3.8E-138 & 7.6E-2 \\
\hline
\end{tabular}
\label{tab:45thF107Comp}
\end{small}
\end{table}

\begin{table}[htbp]
\begin{small}
\centering
\caption{MAPE (\%) of SINet and Related Methods on F10.7 Prediction (at $60^{th}$ Day of Forecast)}
\begin{tabular}{lccccccc}
\hline
 & SINet$_{\mbox{f}}$ & SINet$_{\mbox{r}}$ & ARIMA & LSTM & CNN & LSTM+ & TCN \\
\hline
Fold 1 & \textbf{10.2} & 10.4 & 11.6 & 10.6 & 11.6 & 10.6 & 10.8 \\
$p$-value &  & 2.3E-13 & 6.9E-10 & 1.9E-48 & 1.1E-7 & 4.8E-13 & 5.0E-7 \\
Fold 2 & \textbf{10.0} & 10.7 & 11.6 & 10.8 & 10.5 & 10.6 & 10.5 \\
$p$-value &  & 7.9E-5 & 1.1E-10 & 2.9E-111 & 9.8E-9 & 8.1E-10 & 1.5E-32 \\
Fold 3 & 10.1 & 11.0 & 11.6 & 10.9 & 10.6 & 10.4 & \textbf{10.0} \\
$p$-value &  & 1.3E-11 & 1.0E-10 & 2.7E-71 & 3.3E-8 & 4.6E-8 & 1.0E-11 \\
Fold 4 & \textbf{10.1} & 10.6 & 11.6 & 11.7 & 10.4 & 10.4 & 10.3 \\
$p$-value &  & 9.2E-6 & 6.2E-9 & 2.0E-7 & 1.4E-4 & 1.9E-22 & 4.4E-8 \\
Fold 5 & \textbf{10.1} & 11.3 & 11.6 & 11.7 & 10.4 & 10.4 & 10.4 \\
$p$-value &  & 1.7E-18 & 1.4E-10 & 1.0E-48 & 7.7E-12 & 4.5E-5 & 9.6E-12 \\
\hline
\end{tabular}
\label{tab:60thF107Comp}
\end{small}
\end{table}

\begin{table}[htbp]
\begin{small}
\centering
\caption{MAPE (\%) of SINet and Related Methods on F30 Prediction (at $1^{st}$ Day of Forecast)}
\begin{tabular}{lccccccc}
\hline
 & SINet$_{\mbox{f}}$ & SINet$_{\mbox{r}}$ & ARIMA & LSTM & CNN & LSTM+ & TCN \\
\hline
Fold 1 & \textbf{2.0} & \textbf{2.0} & 2.1 & 2.4 & 3.7 & 2.1 & 2.2 \\
$p$-value &  & --- & 1.2E-2 & 2.0E-78 & 1.9E-299 & 7.8E-26 & 1.6E-44 \\
Fold 2 & \textbf{2.0} & \textbf{2.0} & 2.1 & 2.7 & 3.3 & 2.2 & 2.7 \\
$p$-value &  & --- & 1.1E-1 & 2.7E-127 & 3.4E-235 & 2.5E-23 & 2.9E-146 \\
Fold 3 & \textbf{2.0} & \textbf{2.0} & 2.1 & 2.7 & 3.4 & 2.5 & 2.1 \\
$p$-value &  & --- & 2.5E-2 & 1.0E-149 & 1.1E-279 & 1.0E-96 & 5.9E-23 \\
Fold 4 & \textbf{2.0} & \textbf{2.0} & 2.1 & 2.8 & 3.2 & 2.7 & 2.2 \\
$p$-value &  & --- & 2.0E-1 & 4.0E-138 & 3.4E-200 & 2.0E-126 & 9.2E-35 \\
Fold 5 & \textbf{2.0} & \textbf{2.0} & 2.1 & 2.5 & 3.4 & 2.3 & 2.6 \\
$p$-value &  & --- & 8.2E-2 & 1.4E-73 & 1.4E-270 & 3.0E-48 & 1.5E-108 \\
\hline
\end{tabular}
\label{tab:1stF30Comp}
\end{small}
\end{table}

\begin{table}[htbp]
\begin{small}
\centering
\caption{MAPE (\%) of SINet and Related Methods on F30 Prediction (at $27^{th}$ Day of Forecast)}
\begin{tabular}{lccccccc}
\hline
 & SINet$_{\mbox{f}}$ & SINet$_{\mbox{r}}$ & ARIMA & LSTM & CNN & LSTM+ & TCN \\
\hline
Fold 1 & \textbf{7.1} & 7.9 & 7.7 & 8.4 & 8.2 & 7.8 & 7.4 \\
$p$-value &  & 1.5E-29 & 2.9E-2 & 5.0E-43 & 1.4E-49 & 2.1E-140 & 5.2E-187 \\
Fold 2 & \textbf{7.2} & 7.9 & 7.6 & 8.7 & 8.0 & 7.5 & 7.6 \\
$p$-value &  & 1.0E-25 & 3.8E-2 & 8.1E-65 & 1.2E-62 & 2.3E-65 & 1.5E-265 \\
Fold 3 & \textbf{7.1} & 7.9 & 7.5 & 8.5 & 8.2 & 7.6 & 7.7 \\
$p$-value &  & 2.0E-26 & 5.8E-3 & 3.0E-53 & 2.3E-105 & 8.3E-37 & 1.5E-151 \\
Fold 4 & \textbf{7.1} & 7.9 & 7.5 & 8.5 & 8.3 & 7.6 & 7.3 \\
$p$-value &  & 1.2E-41 & 4.8E-3 & 7.3E-48 & 6.6E-70 & 3.8E-71 & 3.2E-114 \\
Fold 5 & \textbf{7.2} & 7.8 & 7.5 & 8.5 & 8.3 & 7.5 & 7.5 \\
$p$-value &  & 3.8E-52 & 1.4E-2 & 6.2E-40 & 1.3E-64 & 2.0E-51 & 1.5E-65 \\
\hline
\end{tabular}
\label{tab:27thF30Comp}
\end{small}
\end{table}

\begin{table}[htbp]
\begin{small}
\centering
\caption{MAPE (\%) of SINet and Related Methods on F30 Prediction (at $45^{th}$ Day of Forecast)}
\begin{tabular}{lccccccc}
\hline
 & SINet$_{\mbox{f}}$ & SINet$_{\mbox{r}}$ & ARIMA & LSTM & CNN & LSTM+ & TCN \\
\hline
Fold 1 & \textbf{8.7} & 8.8 & 10.6 & 10.4 & \textbf{8.7} & 10.0 & 9.0 \\
$p$-value &  & 1.2E-1 & 6.8E-15 & 4.8E-67 & --- & 1.4E-160 & 2.5E-219 \\
Fold 2 & \textbf{8.7} & 9.0 & 10.6 & 10.6 & 9.0 & 10.2 & 9.2 \\
$p$-value &  & 1.4E-2 & 2.1E-17 & 1.5E-145 & 5.6E-41 & 1.4E-131 & 4.4E-28 \\
Fold 3 & \textbf{8.8} & 8.9 & 10.5 & 10.5 & \textbf{8.8} & 10.0 & 9.0 \\
$p$-value &  & 1.8E-2 & 8.1E-16 & 1.9E-195 & --- & 2.1E-172 & 1.4E-187 \\
Fold 4 & 8.8 & \textbf{8.7} & 10.5 & 10.5 & 9.1 & 9.6 & 9.1 \\
$p$-value &  & 4.8E-1 & 3.1E-17 & 3.5E-76 & 5.1E-54 & 2.2E-113 & 2.5E-46 \\
Fold 5 & \textbf{8.7} & 9.1 & 10.5 & 10.8 & 9.4 & 9.7 & 9.7 \\
$p$-value &  & 5.5E-9 & 1.3E-17 & 4.4E-89 & 2.0E-18 & 2.8E-164 & 3.9E-281 \\
\hline
\end{tabular}
\label{tab:45thF30Comp}
\end{small}
\end{table}

\begin{table}[htbp]
\begin{small}
\centering
\caption{MAPE (\%) of SINet and Related Methods on F30 Prediction (at $60^{th}$ Day of Forecast)}
\begin{tabular}{lccccccc}
\hline
 & SINet$_{\mbox{f}}$ & SINet$_{\mbox{r}}$ & ARIMA & LSTM & CNN & LSTM+ & TCN \\
\hline
Fold 1 & \textbf{9.5} & 9.6 & 10.7 & 10.3 & 10.3 & 9.6 & 9.6 \\
$p$-value &  & 7.4E-32 & 6.1E-10 & 3.3E-85 & 9.6E-37 & 9.0E-25 & 1.4E-11 \\
Fold 2 & \textbf{9.6} & 10.4 & 10.7 & 10.2 & 10.1 & 9.7 & \textbf{9.6} \\
$p$-value &  & 1.6E-9 & 2.5E-8 & 7.9E-71 & 2.2E-35 & 1.1E-20 & --- \\
Fold 3 & 9.5 & 10.2 & 10.7 & 10.2 & 10.2 & 9.6 & \textbf{9.4} \\
$p$-value &  & 4.8E-22 & 8.9E-9 & 1.9E-298 & 9.3E-46 & 4.2E-18 & 8.8E-26 \\
Fold 4 & \textbf{9.5} & 10.1 & 10.7 & 10.5 & 10.1 & 9.6 & 9.6 \\
$p$-value &  & 9.7E-25 & 6.6E-10 & 1.4E-36 & 1.9E-49 & 2.0E-21 & 4.0E-19 \\
Fold 5 & 9.6 & 10.7 & 10.7 & 10.3 & 10.3 & \textbf{9.5} & 9.7 \\
$p$-value &  & 5.8E-45 & 1.5E-9 & 7.9E-117 & 1.1E-35 & 9.6E-21 & 1.2E-5 \\
\hline
\end{tabular}
\label{tab:60thF30Comp}
\end{small}
\end{table}

\section*{Data Availability Statement}

The F10.7 solar index data are available from the NOAA/SWPC F10.7 cm radio emissions archive at
\url{https://www.swpc.noaa.gov/phenomena/f107-cm-radio-emissions}.
The F30 solar index data are available from the daily flux archive
of the Nobeyama Radio Polarimeters at \url{https://solar.nro.nao.ac.jp/norp/data/daily/}.


\end{document}